 \documentclass[prl,reprint,showpacs,showkeys,floatfix]{revtex4-1}
\usepackage[T1]{fontenc}
\usepackage[utf8]{inputenc}
\usepackage{textcomp}
\usepackage{graphicx}
\usepackage{ae,aeguill}
\usepackage{float} 
\usepackage{amsfonts}
\usepackage{sistyle}
\usepackage{amsmath}
\usepackage{xcolor}
\begin{document}
\title{Imaging of Shear Waves Induced by Lorentz Force in Soft Tissues}
\author{P Grasland-Mongrain}
\email{pol.grasland-mongrain@ens-cachan.org}
\author{R Souchon}
\author{F Cartellier}
\author{A Zorgani}
\author{JY Chapelon}
\author{C Lafon}
\author{S Catheline}
\affiliation{Inserm, U1032, LabTau, Lyon, F-69003, France ; Universit\'e de Lyon, Lyon, F-69003, France}
\pacs{*43.20.Gp, *43.20.Jr}
\keywords{shear wave, lorentz force, soft solid, elastography}

\begin{abstract}
This study presents the first observation of elastic shear waves generated in soft solids using a dynamic electromagnetic field. The first and second experiments of this study showed that Lorentz force can induce a displacement in a soft phantom and that this displacement was detectable by an ultrasound scanner using speckle-tracking algorithms. For a 100 mT magnetic field and a 10 ms, 100 mA peak-to-peak electrical burst, the displacement reached a magnitude of 1 $\mu$m. In the third experiment, we showed that Lorentz force can induce shear waves in a phantom. A physical model using electromagnetic and elasticity equations was proposed. Computer simulations were in good agreement with experimental results. The shear waves induced by Lorentz force were used in the last experiment to estimate the elasticity of a swine liver sample.
\end{abstract}

\maketitle
The displacement of a conductor in a magnetic field induces eddy currents. Conversely, the application of an electrical current in a conductor placed in a magnetic field induces a displacement due to Lorentz force \cite{jackson1998classical}. These two phenomena are currently investigated to produce medical images \cite{wen1998hall}. In the technique called Lorentz Force Electrical Impedance Tomography \cite{grasland2013LFEIT}, also known as Magneto-Acoustical Electrical Tomography \cite{xu2005magneto}, an ultrasound beam is focused in a tissue placed in a magnetic field. The displacement of the tissue due to ultrasound in a magnetic field induces an electrical current. The current is measured using electrodes and has been used to produce tissue electrical conductivity interface images. In a ``reverse'' mode, injecting an electrical current in a tissue placed in a magnetic field induces a displacement due to Lorentz force. As in the megahertz range, shear waves decay over a few micrometers, the displacement propagates only through compression waves. These waves can be detected using ultrasound transducers to produce electrical conductivity images. One implementation of this method is called Magneto-Acoustic Tomography with Magnetic Induction \cite{hu2011magnetoacoustic}.

We hypothesized in this study that applying a low frequency (10-1000 Hz) electrical current through a tissue placed in a magnetic field would produce a shear wave within the medium. This could notably have applications in shear wave elastography \cite{muthupillai1995magnetic}, \cite{sandrin2003transient}, \cite{bercoff2004supersonic}, a medical imaging technique used to map the mechanical properties of biological tissues. The mechanical properties of biological tissues are known to be viscoelastic (hence frequency-dependent) \cite{yang2006simple}, \cite{catheline2004measurement}, \cite{nasseri2002viscoelastic}, often anisotropic, e.g. along muscle fibers \cite{gennisson2003transient}, and nonlinear (changing with pre-stress). However, in the field of medical imaging, most applications rely on a simple model, assuming an elastic isotropic linear solid. The viscoelasticity effect has been shown to have only second effect orders \cite{deffieux2009shear} and the synthetic phantoms as used in this study can reasonably be considered as fully isotropic and linear \cite{catheline2003measurement}, \cite{gennisson2007acoustoelasticity}. Under these assumptions, tissue elasticity can be described by two parameters only, e.g. the shear modulus $\mu$ and Poisson's ratio. The shear modulus is related to the shear wave speed $v_s$ and the density of the medium $\rho$ by the equation $\mu= \rho v_s^2$. The linear elasticity of the biological tissue can thus be estimated using this relation by inducing a shear wave in the medium and measuring its speed in each location. Nowadays techniques use an external vibrator \cite{muthupillai1995magnetic}, \cite{sandrin2003transient}, or acoustic radiation force \cite{bercoff2004supersonic} to induce this shear wave. A new way to perform elastography measurements could thus be made possible by demonstrating the ability of the Lorentz force to induce shear waves.

This study was based on four experiments. The first one aimed to show that the Lorentz force could induce a displacement in a soft solid and that this displacement could be detected in ultrasound images. The second experiment proved that the observed displacement was induced by Lorentz force by discarding other potential sources. The purpose of the third experiment was to induce shear waves by Lorentz force and to compare it with the results of a simulation based on a physical model. The last experiment applied the phenomenon of shear waves induced by Lorentz force to perform elastography measurements in a biological tissue.

\begin{figure}[!ht]
	\includegraphics[width=1\linewidth]{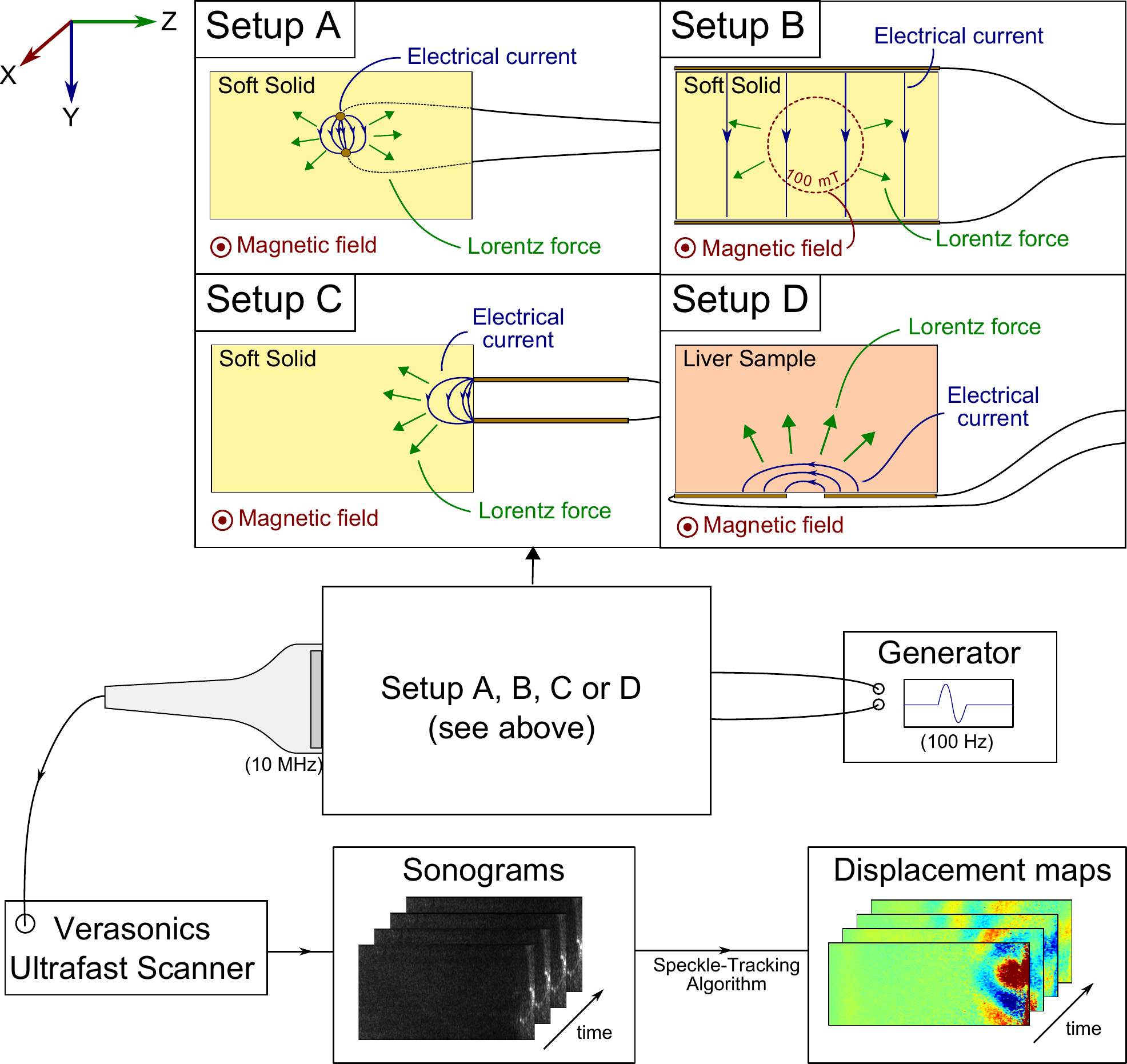}
	\caption{Experimental setup scheme. An electrical current was applied through two planes electrodes in a soft solid placed in a magnetic field. This induced displacements observed with an ultrasound probe.}
	 \label{figElasto1}
\end{figure}

The X axis was defined as the direction of the magnetic field, the Z axis as the main ultrasound propagation axis and the Y axis was placed according to the right-hand rule, as illustrated in Fig. \ref{figElasto1}.

In the experiments, a voltage $\Delta V$ was first applied between two electrodes, leading to an electrical field $\mathbf{E}=-\mathbf{\nabla{V}}$. In a tissue of electrical conductivity $\sigma$ placed between the electrodes, according to Ohm's law the density of current $\mathbf{j}$ was equal to $\sigma \mathbf{E}$. Under the assumptions discussed previously (elastic linear isotropic solid), Navier's equation governed the displacement $\mathbf{u}$ in each point of the tissue submitted to the body force $\mathbf{f}$ \cite{roth1994theoretical}, \cite{aki1980quantitative}:
\begin{equation}
	\rho\frac{d^2\mathbf{u}}{dt^2} = (K + \frac{4}{3}\mu) \nabla (\nabla . \mathbf{u}) + \mu \nabla \times (\nabla \times \mathbf{u}) + \mathbf{f}
	\label{eqElasto1}
\end{equation}
where $\rho$ is the medium density, $\mathbf{u}$ the local displacement, $t$ the time, $K$ the bulk modulus and $\mu$ the shear modulus. In our case, the medium was electrically neutral and its magnetic permeability was close to one \cite{brigadnov2003mathematical}, so we replaced the external force by the Lorentz force $\mathbf{f}=\mathbf{j}\times\mathbf{B}$ where $\mathbf{B}$ is the magnetic field \cite{steigmann2009formulation}.
The displacements were observed in the experiment as illustrated in Fig. \ref{figElasto1}. The components of the apparatus were as follows: a permanent magnet, a conductive medium, two electrodes and an ultrasonic scanner.

The magnetic field was created by a U-shaped magnet with each pole composed by two 3x5x5 cm$^3$ NdFeB magnets, which produced a locally homogenous magnetic field of 300 $\pm$ 50 mT in a volume of 4x4x4 cm$^3$. A single sine cycle of 100 Hz with an amplitude of 80 V peak-to-peak signal was delivered by a generator (AFG 3022B, Tektronix, Beaverton, OR, USA) and an amplifier (A500, Behringer, Willich, Germany). The signal was applied on two wire copper electrodes placed in the middle of a XZ side of a tissue-mimicking phantom. A 1 $\Omega$ resistor was placed between the generator and an electrode. Current flowing through the phantom was estimated by measuring the voltage across the resistor, and was typically between 100 and 200 mA. The medium was a 4x8x8 cm$^3$ water-based phantom made from 5\% polyvinyl alcohol, 0.1 \% graphite powder and 1\% salt. Three freezing/defreezing cycles were applied to stiffen the material \cite{fromageau2003characterization}. The graphite powder was made of submillimeter particles which presented a speckle pattern on the ultrasound images. The medium was observed with a 10 MHz ultrasonic probe made of 128 elements and a Verasonics scanner (Verasonics V-1, Redmond, WA, USA). The probe was used in ultrafast mode \cite{bercoff2004supersonic}, acquiring 1000 ultrasound frames per second. The Z component of the displacement in the medium (``Z-displacement'') was observed by performing cross-correlations between radiofrequency images with a speckle-tracking technique. Although the wavelength was approximately equal to 150 $\mu$m, this technique was capable of measuring displacements under the micrometer size. To remove high frequency noise, displacement images were filtered by setting to zero all frequencies above 200 Hz in the Fourier transform of the displacement over time of each pixel.

This setup, called thereafter Setup A, was designed to have a localized electrical current and a uniform magnetic field.

\begin{figure}[!ht]
	\includegraphics[width=1\linewidth]{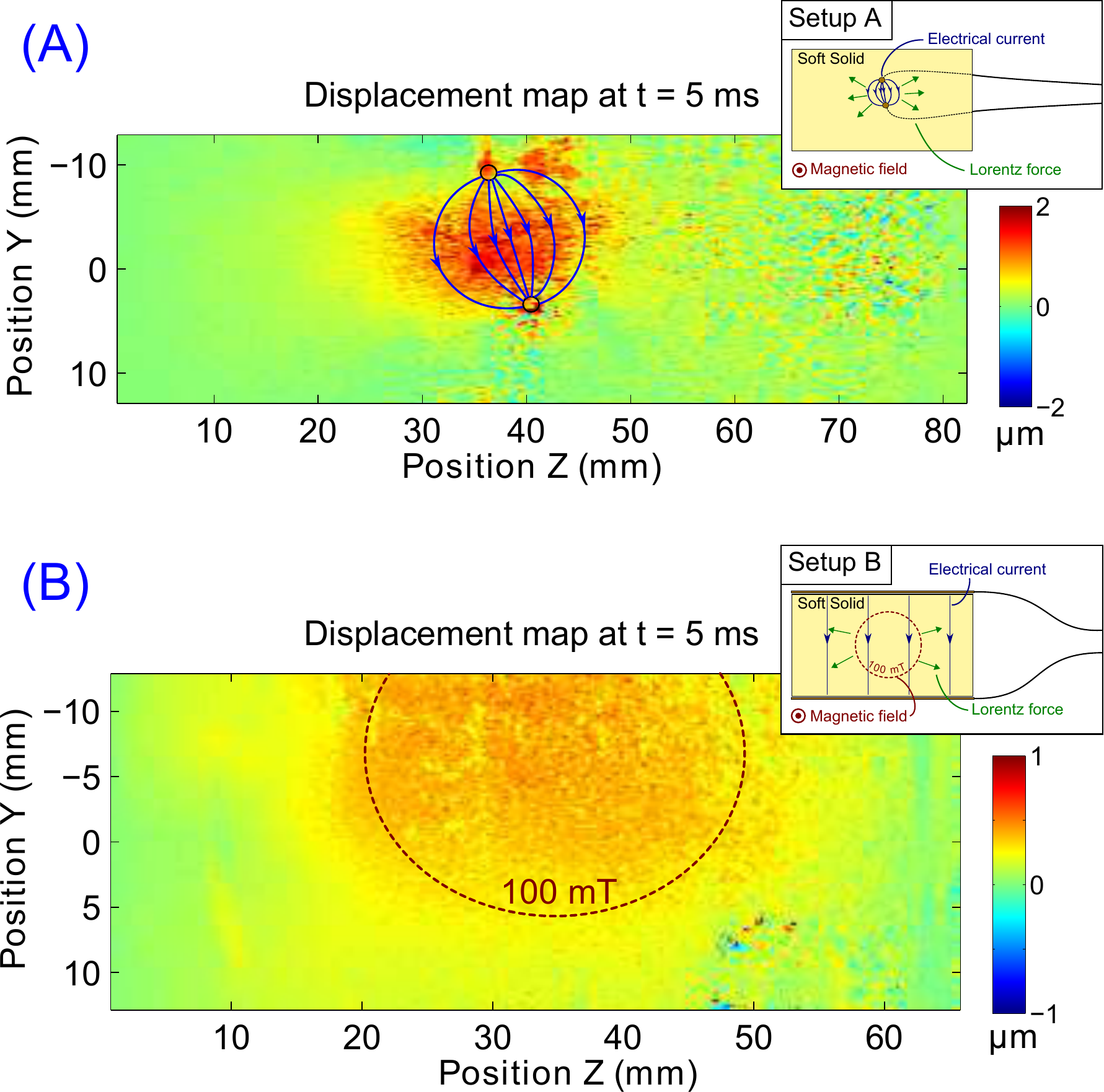}
	\caption{(A) Z-displacement map in the YZ plane in the homogenous phantom with a localized electrical current with Setup A, 5 ms after current injection. The electrodes' locations are circled in black. The expected electrical current lines are drawn with blue lines. Displacement reached an amplitude of 2 $\mu$m in the electrical current location. (B) Z-displacement map in the YZ plane in the homogenous phantom with a localized magnetic field with Setup B, 5 ms after current injection. An isopotential magnetic field lines is drawn in red dots. Displacement reached an amplitude of 1 $\mu$m in the magnetic field location. }
	 \label{figElasto2}
\end{figure}

The Z-displacement observed 5 ms after current injection is illustrated in Fig. \ref{figElasto2}-(A). It is possible to observe a localized displacement occurring between the electrodes, where the current density is highest.

A second setup, the Setup B, was designed to have a uniform electrical current and a localized magnetic field.

The magnet was replaced by a 1x2x1 cm$^3$ NdFeB magnet which produced a 100 $\pm$ 50 mT magnetic field in a 2x2x2 cm$^3$ volume in the phantom. Two flat electrodes were placed on the top and the bottom side of the phantom.

The Z-displacement observed 5 ms after current injection is illustrated in Fig. \ref{figElasto2}-(B). This map shows a 0.5 $\mu$m displacement in presence of magnetic field, proving that this displacement was not due to global motion of the electrodes. Moreover, the root-mean-square of the displacement values was equal to 0.5 $\mu$m, while the same apparatus without any magnetic field gave a root-mean-square value of 0.1 $\mu$m. Amplitude of displacements induced by Lorentz force was consequently at least five times higher than the one due to any other phenomenon like tissue expansion due to heating by Joule effect.
 This experiment thus showed that it was indeed the Lorentz force which induced the observed displacement in a soft solid.

The Setup C was then designed to observe the propagation of displacements as shear waves.

We used the 300 mT magnet and two 4x0.1x10 cm$^3$ planes parallel electrodes, separated by a gap of 1 cm. Their tip was in contact with the medium in order to localize the electrical current near a border. The shear wave propagation can be described by taking the curl of equation (\ref{eqElasto1}), with $\mathbf{s}=\nabla \times \mathbf{u}$:
\begin{equation}
		\frac{ d \mathbf{s}^2}{dt^2} = \frac{\mu}{\rho} \nabla \mathbf{s} + \frac{1}{\rho}\nabla \times (\mathbf{j}\times\mathbf{B})
	\label{eqElasto2}
\end{equation} By using vectorial identity: $\nabla \times (\mathbf{j}\times\mathbf{B}) = (\nabla . \mathbf{B})\mathbf{j} - (\nabla . \mathbf{j})\mathbf{B} + (\mathbf{B}.\nabla)\mathbf{j} - (\mathbf{j}.\nabla)\mathbf{B}$ and the second Maxwell equation $\nabla . \mathbf{B} = 0$ and by noting that $\nabla . \mathbf{j} = 0$ when charges do not accumulate \cite{roth1994theoretical}, we get the equation:
\begin{equation}
		\frac{ d \mathbf{s}^2}{dt^2} = \frac{\mu}{\rho} \Delta \mathbf{s} - \frac{1}{\rho}(\mathbf{B}.\nabla)\mathbf{j} - \frac{1}{\rho}(\mathbf{j}.\nabla)\mathbf{B}
	\label{eqElasto3}
\end{equation} This equation shows that shear motion, described as the curl of the displacement, is created by the variations of $\mathbf{j}$ along magnetic field $\mathbf{B}$ direction (as in Setup A), and equivalently by the variations of $\mathbf{B}$ along density current $\mathbf{j}$ direction (as in Setup B). To model the experiment quantitatively, the Eq. \ref{eqElasto1} was solved using a Green function $G_{kz}(\mathbf{r},t;\mathbf{r}_s,t_s)$ with $\mathbf{r}$ the observation point coordinate, $t$ the time of observation, $\mathbf{r}_s$ position of the source of the body force, $t_s$ time of occurrence and $k$ the direction of the Lorentz force (supposed to be in the YZ plane) \cite{aki1980quantitative}. The Z-displacement $u_z(\mathbf{r},t)$ created by a force $\mathbf{f}$ was calculated using $G_{kz}(\mathbf{r},t;\mathbf{r}_s,t_s)$:
\begin{equation}
	u_z(\mathbf{r},t) = \int_t{ \iiint_V{\mathbf{f}(\mathbf{r}_s,t_s)\otimes G_{kz}(\mathbf{r},t;\mathbf{r}_s,t_s)d^3r_s} dt_s}
	\label{eqElasto4}
\end{equation} Given the complexity of the Green’s function in solids, Eq. \ref{eqElasto4} was computed numerically in a particular case as follows. We modeled two thin parallels electrodes having a respective lineic charge $+\lambda(t)$ and $-\lambda(t)$, of length $2l$ and separated by a distance $2a$. The medium was modeled as an infinite medium with a uniform electrical conductivity and we neglected all boundary effects. The electrical field created by the two electrodes was calculated in a 2D plane. The Z-component of the Lorentz force was then deduced from its expression combined with Ohm's law:
$f_z(y,z) = \sigma B \lambda 
[((\frac{y-a}{l-z})^{2}+1)^{-\frac{1}{2}} + ((\frac{y+a}{l-z})^{2}+1)^{-\frac{1}{2}}  +
((\frac{y-a}{l+z})^{2}+1)^{-\frac{1}{2}} + ((\frac{y+a}{l+z})^{2}+1)^{-\frac{1}{2}}] $.
The following values were used : $2l$ = 0.1 m, $2a$ = 0.01 m, $\sigma B \lambda$ = cos(2$\pi$$\nu t$) with $\nu$ = 100 Hz if 0 < t < 10 ms and $\sigma B \lambda$ = 0 otherwise. The solution of equation (\ref{eqElasto1}) with this expression of the force was then computed numerically, using a compression wave speed of 1480 m.s$^{-1}$ and a shear wave speed of 1.3 m.s$^{-1}$, corresponding to a medium density $\rho$ of 1000 kg.m$^{-3}$ and a bulk modulus $K$ of 2.2 GPa.

\begin{figure*}[!h]
	\includegraphics[width=1\linewidth]{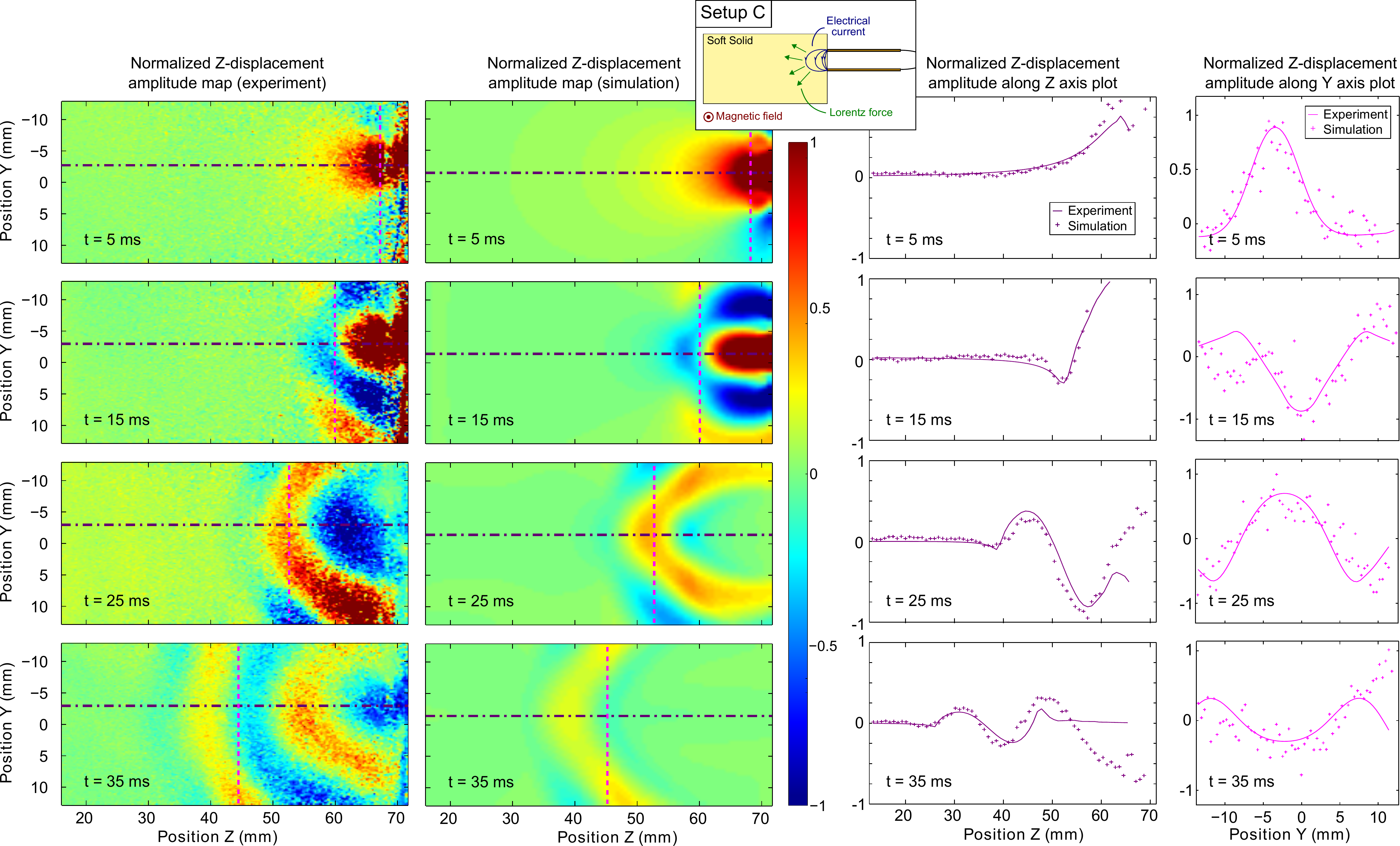}
	\caption{Z-displacement maps in the YZ plane in the homogenous medium with Setup C, respectively 5, 15, 25 and 35 ms after current injection, as given by the experiment and the simulation. Experimental and simulated values are plotted along a Z axis and a Y axis at different depths (respectively 68, 60, 52 and 46 mm away from transducer). Shear waves propagating at 1.3 m.s$^{-1}$ were found. Displacements parallel to the direction of propagation were also observed.}
	 \label{figElasto3}
\end{figure*}

Normalized Z-displacement maps observed 5, 15, 25 and 35 ms after current injection as given by the experiment and the simulation are illustrated in Fig. \ref{figElasto3}. To have a more quantitative comparison, the experiment and simulation results along two lines are plotted, the first along the Z axis between the two electrodes and the second along the Y axis respectively 2, 10, 18 and 26 mm away from the electrodes. The maximum displacement found by the experiment was equal to 2 $\mu$m. The displacement propagation speed was equal to 1.3 m.s$^{-1}$ and was in line with reported shear wave speeds in soft solid. Displacements parallel to the direction of propagation were also observed, which is a typical near field feature \cite{catheline1999solution}. Differences in the waves trail observable at time $t$ = 35 ms were interpreted as the consequence of the approximation of infinite medium in the Green function calculation which did not take into account any boundary effect of the medium, especially the rigid contact with the electrodes. Good general agreement was however observed between experiment and simulation, qualitatively and quantitatively. 

The purpose of the Setup D was to induce shear waves due to Lorentz force in a biological sample.

The sample was a 3x3x3 cm$^3$ lobe of swine liver cube, unfrozen in a 1 \% salt at 20$^\mathrm{o}$C solution. As the ultrasound attenuation of this tissue was high, displacement was difficult to compute farther than one centimeter from the transducer. The two electrodes were consequently positioned under the sample, separated by a gap of 1 cm, in order to see the propagation of shear waves along the Y axis over a few centimeters.

\begin{figure}[!ht]
	\includegraphics[width=1\linewidth]{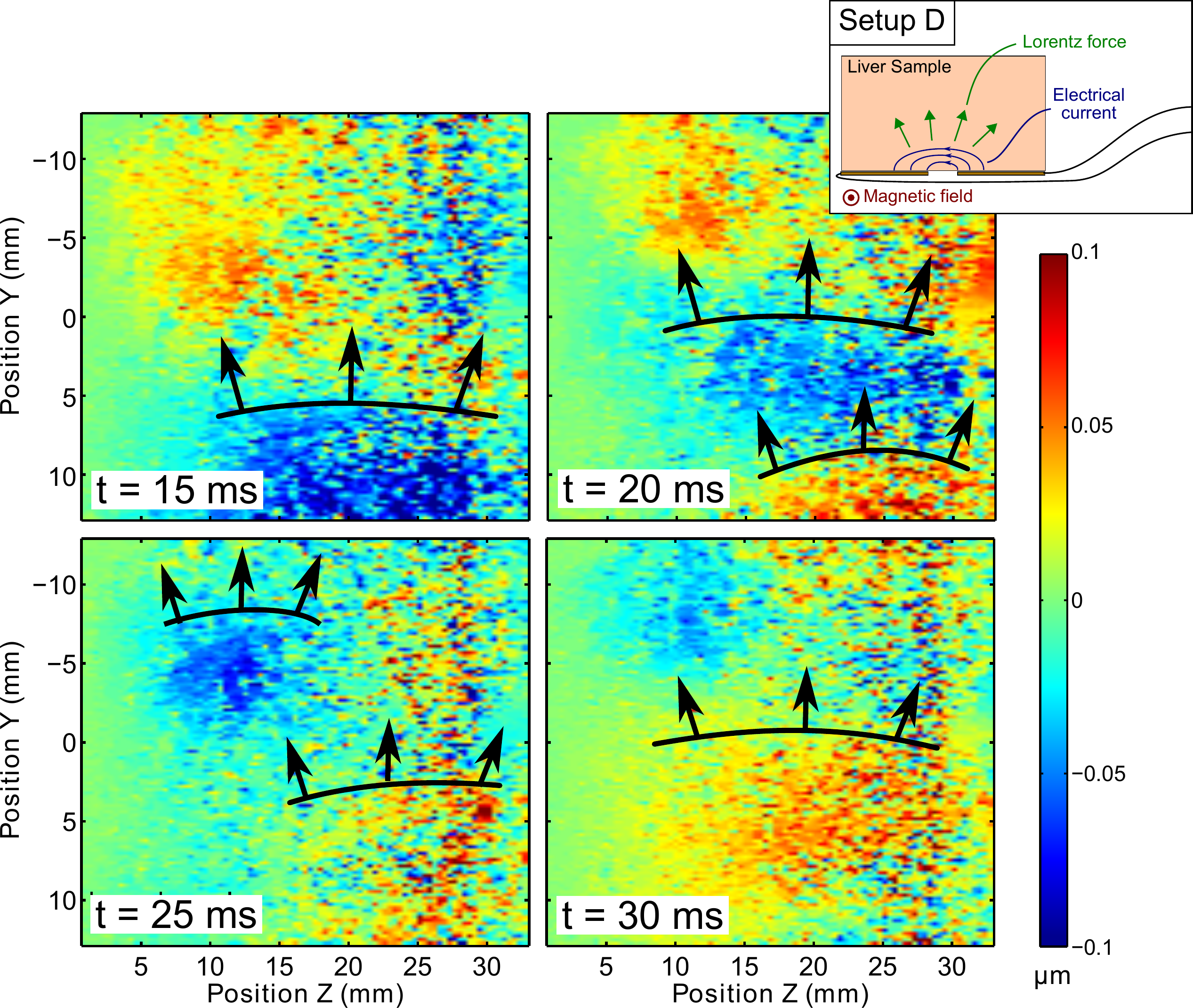}
	\caption{Z-displacement map in the YZ plane in the liver medium with Setup D, respectively 15, 20, 25 and 30 ms after current injection.}
	 \label{figElasto4}
\end{figure}

Z-displacement maps observed 15, 20, 25 and 30 ms after current injection are illustrated in Fig. \ref{figElasto4}. The maximum displacement was equal to 0.2 $\mu$m. Shear waves propagated at a velocity of 1.4 $\pm$ 0.2 m.s$^{-1}$. Estimating the elasticity with $\mu = \rho v_s^2$ relationship gave a shear modulus of 2.0 $\pm$ 0.6 kPa. This value is in excellent agreement with in-vivo measurements in healthy human liver with magnetic resonance elastography measurements \cite{rouviere2006mr}. This indicated that the technique has the potential to be applied in biological tissues.

These four experiments demonstrated that (1) Lorentz force can induce displacements in soft tissues and these displacements are observable with an ultrasound scanner, (2) the Lorentz force can be used to generate shear waves, and (3) this phenomenon can be used to perform shear wave elastography measurements in biological tissues.

The typical displacement was at the micrometer scale, close to the sensitivity of magnetic resonance elastography \cite{muthupillai1995magnetic} and ultrasound elastography \cite{nightingale2002acoustic}. The electrical current, estimated at 100 mA, was about a hundred time higher than the highest current tolerated in the human body according to international standards \cite{iec60479-1}. However, the magnetic field strength of a magnetic resonance imaging system, ten times higher than the one used here, could increase the observed amplitude. One particular advantage of this principle lies in the possibility of inducing displacement remotely, either by applying directly the electrical current by electrodes or by inducing it with a time-varying magnetic field. Another advantage lies in the possibility to choose precisely the shear wave source frequency, closely related to the emitted current frequency. It would have interest for studying the properties of the soft solids at different frequencies.

Additional materials (Z-displacement maps videos of the experiments) are available on-line.

\bibliography{biblio}
\end{document}